\documentclass[showpacs,amsmath,amssymb,prl,twocolumn]{revtex4-1}

\usepackage{graphicx}% Include figure files
\usepackage{dcolumn}% Align table columns on decimal point
\usepackage{bm}% bold math

\begin{document}

\preprint{APS/123-QED}

\title{Full statistics of ideal homodyne detection  using real (noisy) local oscillator.}

\author{A. Auyuanet, E. Benech, H. Failache and A. Lezama}
\email{alezama@fing.edu.uy}
\affiliation{Instituto de F\'{\i}sica, Facultad de Ingenier\'{\i}a,
Universidad de la Rep\'{u}blica,\\ J. Herrera y Reissig 565, 11300
Montevideo, Uruguay}%

\date{\today}

\begin{abstract}
We show that the full statistics of the two detectors outputs in a balanced homodyne detection setup involving a local oscillator in an ideal coherent state is  experimentally accessible despite the excess noise existing in actual laser sources. This possibility is illustrated using  phase randomized coherent states signals from which the statistics of Fock states can  accurately be obtained. The experimental verification of the recently predicted [K\"{u}hn and Vogel, Phys. Rev. A, \textbf{98}, 013832 (2018)] two-detector correlation probability for Fock states $\vert 1 \rangle$ and $\vert 2 \rangle$ is presented.
\end{abstract}

\maketitle

\section{Introduction}

An important task in the field of quantum optics is the development of theoretical and experimental tools to identify  the nonclassical features of the light states that can be prepared in actual experiments. 

One of the most successful techniques so far employed is the balanced homodyne  detection (BHD) where the signal field  under investigation is mixed with a strong local oscillator (LO) in a balanced beamsplitter (BS) and two detectors are used to measure light intensity at the BS output ports. The photocurrent outputs of the two detectors are electronically subtracted. This technique allows the measurement of field quadratures, a continuous variable observable \cite{Yuen83,Vogel93,Raymer95,Leonhardt97}, and plays an important role in quantum information processing. It has been used to observe light squeezing \cite{Slusher87,Ladd,Lvovsky15}, to generate non-classical states \cite{Ourjoumtsev}, to perform quantum state tomography \cite{Smithey}  and  quantum teleportation \cite{Furusawa706}. 

One of the key advantages of the use of the BHD technique for quadrature measurement comes from its insensitivity to noise in the LO. The fluctuations produced by excess noise in the LO amplitude are equally present in the two detectors and consequently canceled in the electronic subtraction. The cost of BHD insensitivity to noise is the loss of information since more information is present at the two detectors outputs that remains in their difference. 

Several recent articles have addressed  the possibility of obtaining additional information on the signal light state from the  the full statistics of the two detector outputs in homodyne detection using balanced and unbalanced BS \cite{Vogel95,Wallentowitz96,Kuhn17}. Recently,  the probability distribution for the product of the photodetectors outputs (correlation) for several states of the light field has been calculated and   nonclassicality criteria were established based in the correlation distribution alone \cite{kuhn18}. 

In general, these approaches rely on the use of a LO in a coherent state whose noise level is limited to shot noise. Such ideal LO is seldom available in laboratories were actual laser sources generally present excess noise substantially above the shot noise limit.

In this article we show that for the BHD scheme, the complete photocurrent statistics, as would be obtained with a coherent state LO, is experimentally accessible despite the LO excess noise at the relatively low cost of an additional measurements set. 

We illustrate this possibility by obtaining first from experimental observations the  ideal two-detectors BHD statistics of  phase randomized coherent states (PRCS). 

PRCS have been used in quantum key distribution (QKD) protocols \cite{Scarani09,Jouguet13} as substitutes  for single-photon pulses \cite{Hwang03,Lo05,Zhao06,Aragoneses18}. Extending the security analysis of these QKD implementations \cite{Yuan16}, we have recently shown that PRCS can be used to test and simulate single-photon performances in arbitrary linear optics processes \cite{Valente17}. Applying this approach to the BHD process, we have measured the complete two-detector statistics and the correlation probability distribution for Fock states recently predicted in \cite{kuhn18}.

\section{Background}

\begin{figure}[htb]
\centering\includegraphics[width=7cm]{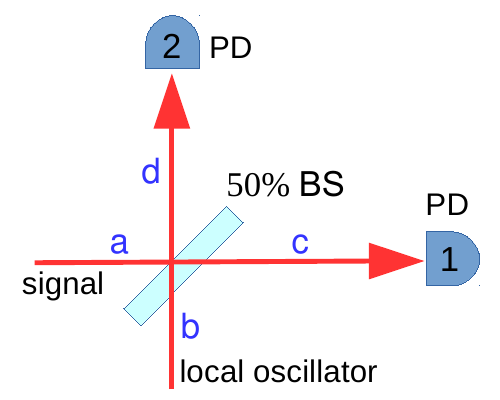}
\caption{\label{BS} Balanced homodyne detection arrangement (BS: beamsplitter, PD: photodetector). } 
\end{figure}

The typical scheme for balanced homodyne detection is reminded in Fig. \ref{BS}. A 50\% beamsplitter (BS) combines two modes of the electromagnetic field with the same frequency. The signal field is incident on one of the input ports of the BS while a strong field, the local oscillator, is incident on the other input port. The corresponding annihilation  operators are $a$ and $b$ respectively. The annihilation  operators $c$ and $d$ correspond to the field at the output of BS. They are related to the incident field operators through the BS transformation which can be written as:
\begin{eqnarray}
\left(\begin{matrix}
 c\\ d 
\end{matrix}\right) = \frac{1}{\sqrt{2}}\left(\begin{matrix}
  1 & 1\\-1 & 1 
\end{matrix}\right) \left(\begin{matrix}
  a\\ b 
\end{matrix}\right)\label{bst}
\end{eqnarray}
where a possible global phase factor has been ignored. 

We are concerned with the field fluctuations: $\delta a=a -\langle a\rangle$ and $\delta b=b-\langle b\rangle$. These operators verify the usual bosonic commutation rules: $[\delta a,\delta a]=0$, $[\delta a,\delta a^{\dagger}]=1$ and similarly for $\delta b$. We concentrate in states of the field $a$ for which $\langle a\rangle = 0$. Let $\alpha$ designate the mean value of the LO ($\langle b\rangle = \alpha$). Since the LO oscillator is the phase reference, $\alpha$ can be taken real with no loss of generality. We assume that the LO is intense and that its fluctuations are small compared to the mean value: $\vert \alpha\vert^2\gg 1,  \langle \delta a^{\dagger}\delta a \rangle,  \langle \delta b^{\dagger}\delta b \rangle$. 

The two detectors $1$ and $2$ collect the light from the BS output ports. We assume perfect detection efficiency (we will return to this assumption later) and neglect dark currents (a safe assumption for sufficiently strong LO).  The two photodetectors output are a measure of the photon number operators of field modes  $c$ and $d$ respectively. Using the BS transformation (\ref{bst}) and neglecting the terms independent of $\alpha$, we obtain for the photocurrent fluctuations:
\begin{subequations}\label{Ifluct}
\begin{eqnarray}
\delta I_1 &=& \frac{1}{2}[\alpha(\delta a+\delta a^{\dagger})+\alpha(\delta b+\delta b^{\dagger})]\\
\delta I_2 &=& \frac{1}{2}[-\alpha(\delta a+\delta a^{\dagger})+\alpha(\delta b+\delta b^{\dagger})]
\end{eqnarray}
\end{subequations}

We are interested in the joint statistics of the outcome of the measure of $\delta I_1$ and $\delta I_2$. Since the two observables are related to both fields, in general their outcomes are correlated. 
%It is the purpose of this article to explore these correlations.

It is convenient to introduce the operators corresponding to the sum and difference of  photocurrent fluctuations: $\delta S=\delta I_1+\delta I_2$, $\delta D=\delta I_1-\delta I_2$. We have:
\begin{subequations}\label{SyD}
\begin{eqnarray}
\delta D &=& \alpha(\delta a+\delta a^{\dagger})\\
\delta S &=& \alpha(\delta b+\delta b^{\dagger})
\end{eqnarray}
\end{subequations}
$\delta D$ and $\delta S$ are proportional to the amplitude quadrature fluctuations $( \delta a+\delta a^{\dagger})$ and  $( \delta b+ \delta b^{\dagger})$ respectively. Since $\delta D$ and $\delta S$ just depend on one field, they represent independent  random variables \cite{noentanglement}. Let $P_D(x)$ and $P_S(x)$ be the corresponding probability densities. 

If the field $a(b)$ is in a coherent state, the fluctuations  $\delta D(\delta S)$ are Gaussian with variance $\sigma_0^2=\alpha^2$. These fluctuation define the shot-noise level.\\

%The corresponding probability distribution is:
%
%\begin{eqnarray}Experimental setup. FBS: fiber beam splitter, FPM: fiber phase modulator, PBS: polarization beam splitter, NDF: neutral density filters, HWP: half-wave plate, PD: photodetector.
%P[\delta D(\delta S)=x] = \frac{1}{\sqrt{2\pi\sigma_0^2}}e^{-\dfrac{x^2}{2\sigma_0^2}}
% \label{Gaussiano}
%\end{eqnarray}
%where  is the variance of the vacuum fluctuations representing the shot noise level. 

Let $P(x,y)$ be the probability distribution for the outcomes $\delta I_1=x$ and $\delta I_2=y$. From the stochastic independence of $\delta D$ and $\delta S$ it results:

\begin{eqnarray}
P(x,y)&=& 2P_S(x+y)P_D(x-y)  \label{joint}
\end{eqnarray}\\

\subsection{Correlation statistics}

An important characterization of the two detector statistics is given by the  probability  distribution  $w(M)$ of the product of events at the two photodetector outputs ($M\equiv\delta I_1\delta I_2$) \cite{kuhn18}.  Here again it is suitable to relate the random variable $M$ to the stochastically independent  $\delta D$ and $\delta S$. Since $M=\frac{1}{4}(\delta S^2-\delta D^2)$ we have:
%\begin{eqnarray}
%M=\frac{1}{4}(\delta S^2-\delta D^2)\label{MSD}
%\end{eqnarray}
%from which we derive:
\begin{eqnarray}
w(M)&=& 4\int_{-\infty}^{\infty} Q_{S^2}(4M+v)Q_{D^2}(v) dv \label{wM}
\end{eqnarray}
where $Q_{D^2}(x)$ and $Q_{S^2}(x)$ are the probability densities for $\delta D^2$ and $\delta S^2$ respectively. 
\begin{eqnarray}
Q_{D^2}(x)&=& \dfrac{P_D(\sqrt{x})}{\sqrt{x}}\Theta(x) \label{probCuad}
\end{eqnarray}
and similarly for $Q_{S^2}(x)$. Here $\Theta(x)$ is the Heaviside function ($\Theta(x)=0$ for $x<0$ and $\Theta(x)=1$ for $x>0$).

\subsection{Non-ideal homodyne detection}

In actual experiments the local oscillator is not in a well defined coherent state. While some sophisticated laser sources  have amplitude noise levels approaching that of a coherent state \cite{Seifert06}, most lasers present amplitude noise well above the  shot noise level. As an example, the extended cavity CW diode laser used in our experiments shows $\sim 30$ dB excess noise in the few MHz frequency range. Such excess noise largely dominates the fluctuations of $\delta S$ (Eq. \ref{SyD}b) and introduces a strong correlation between the photodetectors outputs $\delta I_1$ and $\delta I_2$. In turn, such correlation results in a strong bias of the distribution $w(M)$ towards positive values of $M$ as expected for purely classical fluctuations. As discussed in \cite{kuhn18}, many interesting features of $w(M)$ for nonclassical light states are revealed for negative values of $M$. For instance Fock states have  negative mean values of  $M$ indicating anti-correlation a feature that is reminiscent of the Hong-Ou-Mandel 
effect. Also, the positive mean value of $M$ observed using a noisy local oscillator is just an indication of classical correlations and not of squeezing as it would be when the LO is in a coherent state \cite{kuhn18}.\\

In view of the detrimental effects of the LO excess noise two questions can be raised: \textit{a}) is it possible in spite of the excess noise to detect nonclassical features in joint statistics of the photodetectors outputs? and \textit{b}) is it possible to extract from the experimental observations the full statistics corresponding to the ideal measurement using a local oscillator in a coherent state?

The answer to both questions is affirmative as a consequence of the fact that $\delta D$ is unaffected by the LO excess noise (Eq. \ref{SyD}a). 
%This well known fact is at the essence of the  homodyne balanced detection technique, the standard method for field quadrature measurement. 
It has long been established that the field nonclassicality can be directly revealed from the quadrature probability distribution through the Vogel criterion \cite{Vogel00,Lvovsky02,Richter02}: if for some frequency component the absolute value of the Fourier transform of the quadrature probability density exceeds the corresponding value for the vacuum, then the state is nonclassical. 
%The complete statistics of the two-detector fluctuations in homodyne measurement obviously contains more information than one  quadrature noise statistics. However the latter is sufficient to establish nonclassicality. 
In the presence of excess noise in the LO, the joint probability distribution given in Eq. (\ref{joint}) is stretched along the main diagonal ($x=y$) of the $x, y$ plane by the classical correlations. However the distribution along the $x=-y$ diagonal is unaffected by the excess noise and represents, up to a factor, the quadrature probability density.

The answer to question \textit{b)} is a consequence of the statistical independence of $\delta D$ and $\delta S$. In order to obtain the ideal statistical properties (LO in a coherent state) all that is required is the knowledge of $P_S(x)$ for a coherent state local oscillator. This probability distribution is the same as $P_D(x)$ for a coherent state or the vacuum (see Eqs. \ref{SyD}) and can be readily obtained experimentally at the only cost of an additional measurement set. Actually this distribution is Gaussian and entirely determined by its variance. In consequence, two independent sets of measurements  suffice for the reconstruction of the full ideal statistics corresponding to the ideal (coherent state) local oscillator: a first set of quadrature measurements with a blocked input port of the BS aimed to determine the variance $\sigma_0^2$ of vacuum fluctuations and a second set of measurements  of the quadrature ($\delta D$) of the state under investigation. With this information at hand, Eqs. \ref{joint} and \ref{wM} become for the ideal LO:

\begin{eqnarray}
P_0(x,y)&=& \dfrac{2}{\sqrt{2\pi\sigma_0^2}}\exp{\left[ -\dfrac{(x+y)^2}{2\sigma_0^2}\right] }P_D(x-y)\label{newjoint} \\  
w_0(M)&=& \frac{8}{\sqrt{2\pi\sigma_0^2}}\int_{\eta}^{\infty}\dfrac{e^{ -\frac{4M+x^2}{2\sigma_0^2}} }{\sqrt{4M+x^2}}P_D(x) dx \label{w}
\end{eqnarray}
with $\eta=\sqrt{Max(0,-4M)}$.\\
%A full discussion of the correlation statistics $w_0(M)$ including the effects of detector efficiency and dark current can be found in \cite{kuhn18} together with the corresponding plots for a variety of quantum states of the field.\\

The Vogel criterion establishes that the knowledge of the quadrature probability distribution and the variance of vacuum fluctuations are sufficient to establish the nonclassicality of the light state.  Eqs. (\ref{newjoint}) and (\ref{w})  show that the same information  also allows the  complete determination of the full joint statistics of the two photodetectors. 

In a recent article, Park et al. \cite{Park17} have presented an alternate method for the identification of nonclassical states. In their approach $P_0(x,y)$ given by Eq. (\ref{newjoint}) is considered as a ``fictitious Wigner function" for which nonclassicality criteria can be established.

\section{Experiment}

\begin{figure}[h]
\centering
{\includegraphics[width=0.7\linewidth]{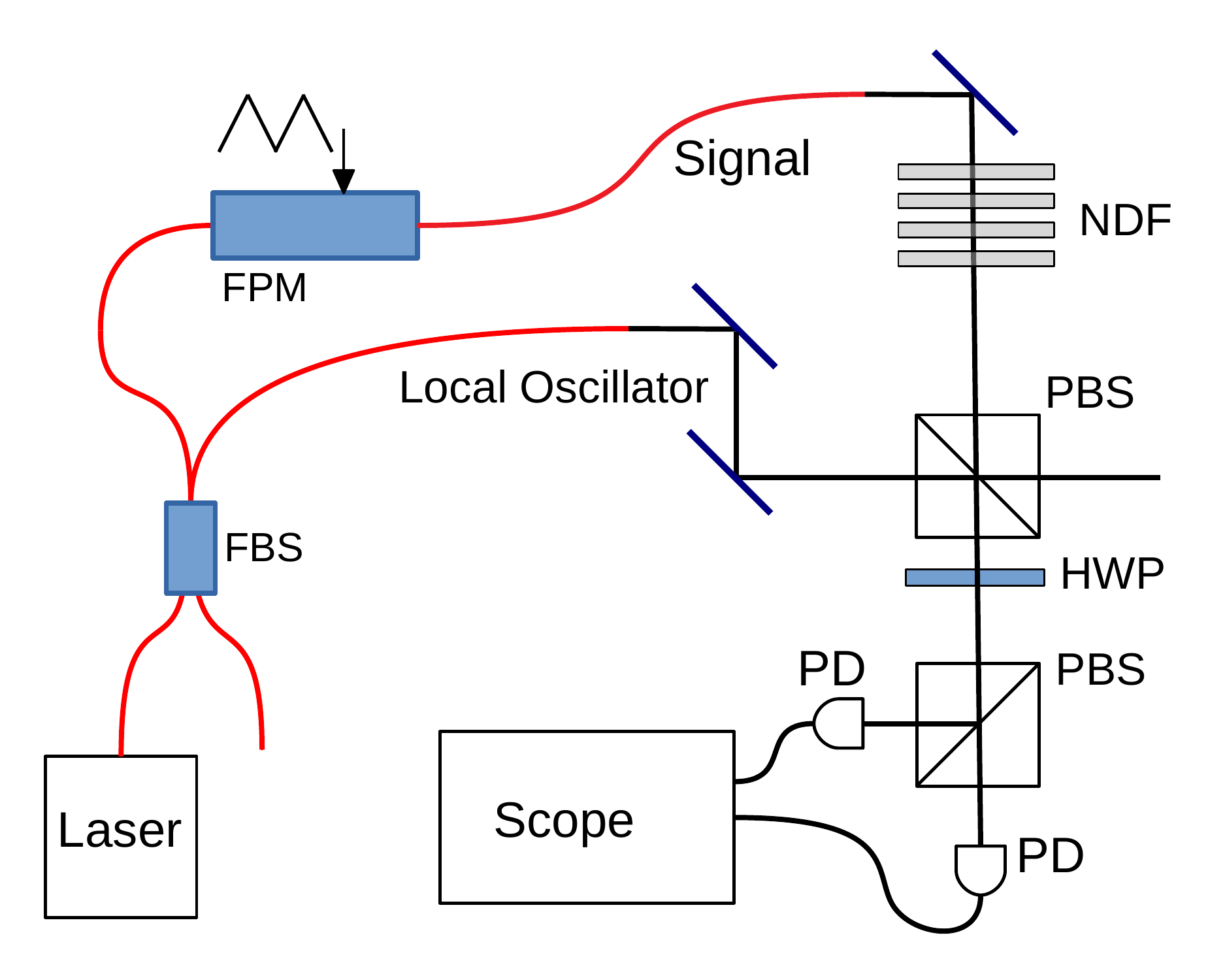}}
\caption{Experimental setup. FBS: fiber beam splitter, FPM: fiber phase modulator, PBS: polarization beam splitter, NDF: neutral density filters, HWP: half-wave plate, PD: photodetector.}
\label{setup}
\end{figure}

The experimental scheme is shown in Fig. \ref{setup}. We have used as  local oscillator the light from of an extended cavity CW diode laser operating at 795 nm and sent through a single mode optical fiber for mode shaping. 
%Fig. \ref{ruido} shows the power spectrum of the sum and difference of the photocurrent fluctuations with only the LO incident on the BS (vacuum in the input port) in the range 0-10 MHz. The noise power for the sum exceeds by ** dB the shot noise level given by the photocurrent difference power spectrum. 

The BHD setup is implemented using two polarization beamsplitters (PBS). The signal and the LO beams are spatially overlapped in a first PBS. A half-wave plate is used to rotate the two orthogonal polarizations of the signal and the LO so that they project equally on the two output polarizations of the second PBS. This arrangement allows a  fine control of the intensity balance of the light intensity on the two photodetectors. The total power at the detectors is 11 mW.

We consider temporal modes of the field with constant amplitude during 400 ns time intervals. For each time interval simultaneous photocurrent output measurements are carried on the two detectors and the results of $10^6$ consecutive  measurement pairs recorded in a fast digital oscilloscope. 

The signal field used in the experiments corresponds to a highly attenuated phase-randomized fraction of the laser light used for the LO. Phase randomization is achieved during the acquisition time with a fiber coupled phase-modulator submitted to a 100 KHz triangular voltage ramp corresponding to  $6\pi$ total phase excursion. The effect of the phase randomization on the signal field quadrature fluctuations, which are comparable in magnitude to the field amplitude, largely exceeds the effect of the laser amplitude fluctuations (which are typically three order of magnitude smaller than the field mean amplitude). In consequence the prepared state of the signal field corresponds to a good approximation to a phase-randomized coherent state (PRCS). 

Phase randomized coherent states correspond to statistical mixtures of Fock states. The density matrix $\rho_{\mu}$ of a PRCS with mean photon number $\mu$ is:
\begin{eqnarray}
\rho_{\mu}\equiv \frac{1}{2\pi}\int_{0}^{2\pi}\vert \sqrt{\mu}e^{i\phi}\rangle\langle \sqrt{\mu}e^{i\phi}\vert d\phi&=& e^{-\mu}\sum_{n=0}^{\infty}\dfrac{\mu^n}{n!}\vert n\rangle\langle n\vert \label{ro}
\end{eqnarray}

The mean photon number $\mu$ corresponding to an experimental PRCS signal  can be accurately determined by fitting the corresponding histogram of the photocurrent difference to the theoretical marginal probability distribution of a PRCS \cite{Valente17}.  

 \begin{figure}[h!]
\centering\includegraphics[width=0.7\linewidth]{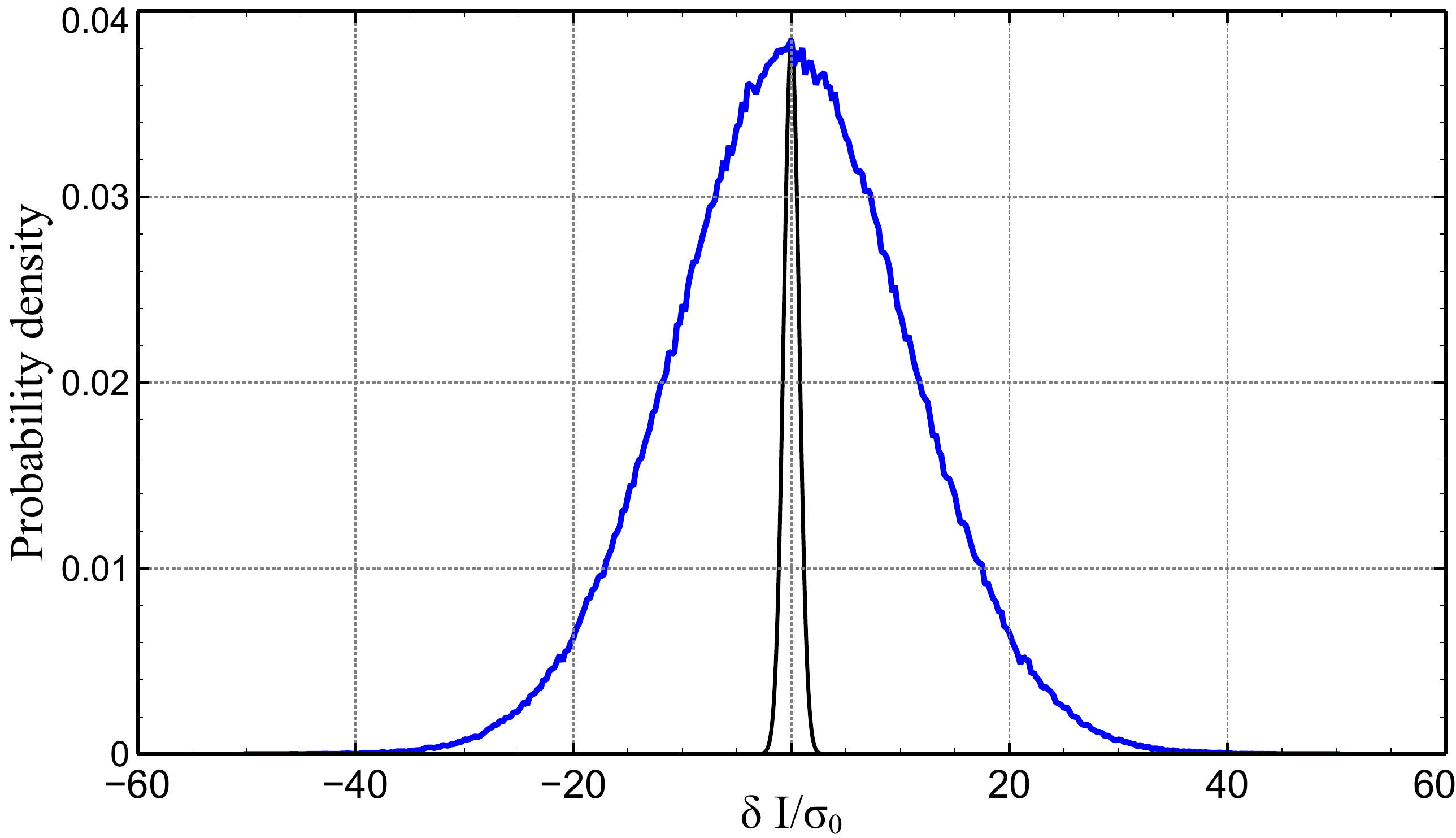}
\caption{\label{ruido} Color online. Blue: Normalized histogram of photocurrent fluctuations for one detector. Black: shot noise contribution (not normalized).} 
\end{figure}

Figure \ref{ruido} shows the fluctuation histogram of the output of one of the two  photodetectors when the signal field is blocked (vacuum state). In all plots in this paper the photodetector outputs are given in units of $\sigma_0$ (the vacuum field quadrature standard deviation).  Also shown in this figure is the corresponding contribution from vacuum fluctuations. As observed,  the histogram is largely dominated by the laser excess noise (26 dB above shot noise level). 

 \begin{figure}[htb]
\centering\includegraphics[width=0.8\linewidth]{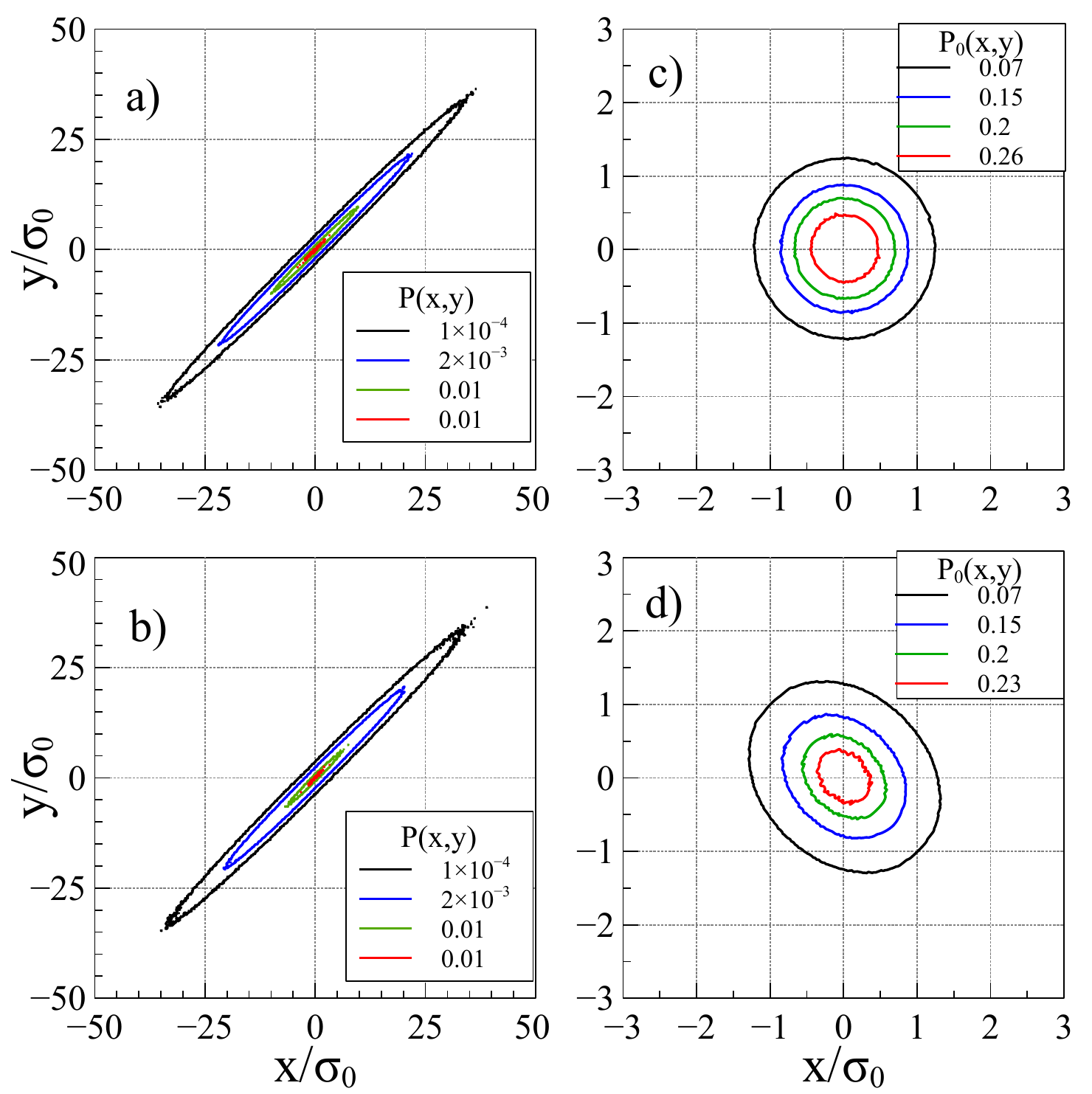}
\caption{\label{pc1c2} Color online. Contour map of the joint probability distribution of the two photodetector outcomes. a) and c) correspond to a vacuum input state. b) and d) correspond to a PRCS with $\mu=0.25$. a) and b) are the  observed statistical distributions with the LO actually used in the experiment. c) and d) correspond to the probability distribution that would be obtained with an ideal LO in a coherent state according to Eq. (\ref{newjoint}) } 
\end{figure}

Figure \ref{pc1c2} shows the two-dimensional contour map representation  of the observed joint statistical distribution of the two photodetectors outputs. Fig. \ref{pc1c2}a corresponds to an incident field in the vacuum state and  \ref{pc1c2}b to a PRCS with mean photon number $\mu=0.25$. The large concentration of the outcomes along the main diagonal ($\delta I_1 \simeq \delta I_2$) is the consequence of the large classical correlation. Figures \ref{pc1c2}c and \ref{pc1c2}d represent the probability distributions corresponding to an ideal local oscillator in a coherent state obtained using Eq. (\ref{newjoint}). As expected, the vacuum state has a rotationally symmetric probability distribution (Gaussian) while the PRCS shows an asymmetry revealing anticorrelation of the photodetector outputs. This anticorrelation  is expected for PRCS as a consequence of energy conservation in the destructing/constructing interference occurring at the two output ports of the BS.\\

\subsection{Nonclassical light states}

The PRCS signal states prepared in our experiment are obviously classical according to the usually  accepted criterion based in the Glauber-Sudarshan P representation: they are statistical mixtures of coherent states. However, they can be used for the experimental characterization of photon number states in an arbitrary \emph{linear} quantum process as has been shown in \cite{Valente17}. The method is briefly outlined here:

Consider a quantum $\rho_{out}=\mathcal{L}(\rho_{in})$ process where $\mathcal{L}(\cdot)$ represents a (generally non-unitary) linear transformation. As a consequence of linearity, from Eq. (\ref{ro}) we have: 
\begin{eqnarray}
\mathcal{L}(\rho_{\mu})&=& e^{-\mu}\sum_{n=0}^{\infty}\dfrac{\mu^n}{n!}\mathcal{L}(\vert n\rangle\langle n\vert) \label{linearity}
\end{eqnarray}

If a linear quantum process is tested for a sufficient number of PRCSs with different values of $\mu$, Eq. (\ref{linearity}) can be inverted and the result of the quantum process on individual Fock states obtained with good approximation \cite{Yuan16}. The homodyne detection setup is a particular example of linear quantum process for which the method in \cite{Yuan16,Valente17} can readily be applied.

If only the vacuum and one PRCS with nonzero $\mu$ are used, a good approximation to the single-photon yield is given by: 
\begin{eqnarray}
\mathcal{L}_1&\simeq & \dfrac{e^{\mu}\mathcal{L}(\rho_{\mu})-\mathcal{L}_0}{\mu} \label{conunsolomu}
\end{eqnarray}
where $\mathcal{L}_n\equiv\mathcal{L}(\vert n\rangle\langle n\vert)$. If two nonzero values of $\mu$ are available, approximations for $\mathcal{L}_1$ and $\mathcal{L}_2$ are simultaneously obtained.
\begin{subequations} \label{masdeunmu}
\begin{eqnarray}
\mathcal{L}_1&\simeq & \dfrac{A_1\mu_2^2-A_2\mu_1^2}{2\Delta} \label{1F}\\
\mathcal{L}_2&\simeq & \dfrac{A_2\mu_1-A_1\mu_2}{\Delta} \label{2F}
\end{eqnarray}
\end{subequations}
Here $A_i\equiv e^{\mu_i}\mathcal{L}(\rho_{\mu_i})-\mathcal{L}_0 \nonumber$ and $\Delta \equiv \frac{1}{2} (\mu_1\mu_2^2-\mu_2\mu_1^2)$. 

 \begin{figure}[h!]
\centering\includegraphics[width=0.8\linewidth]{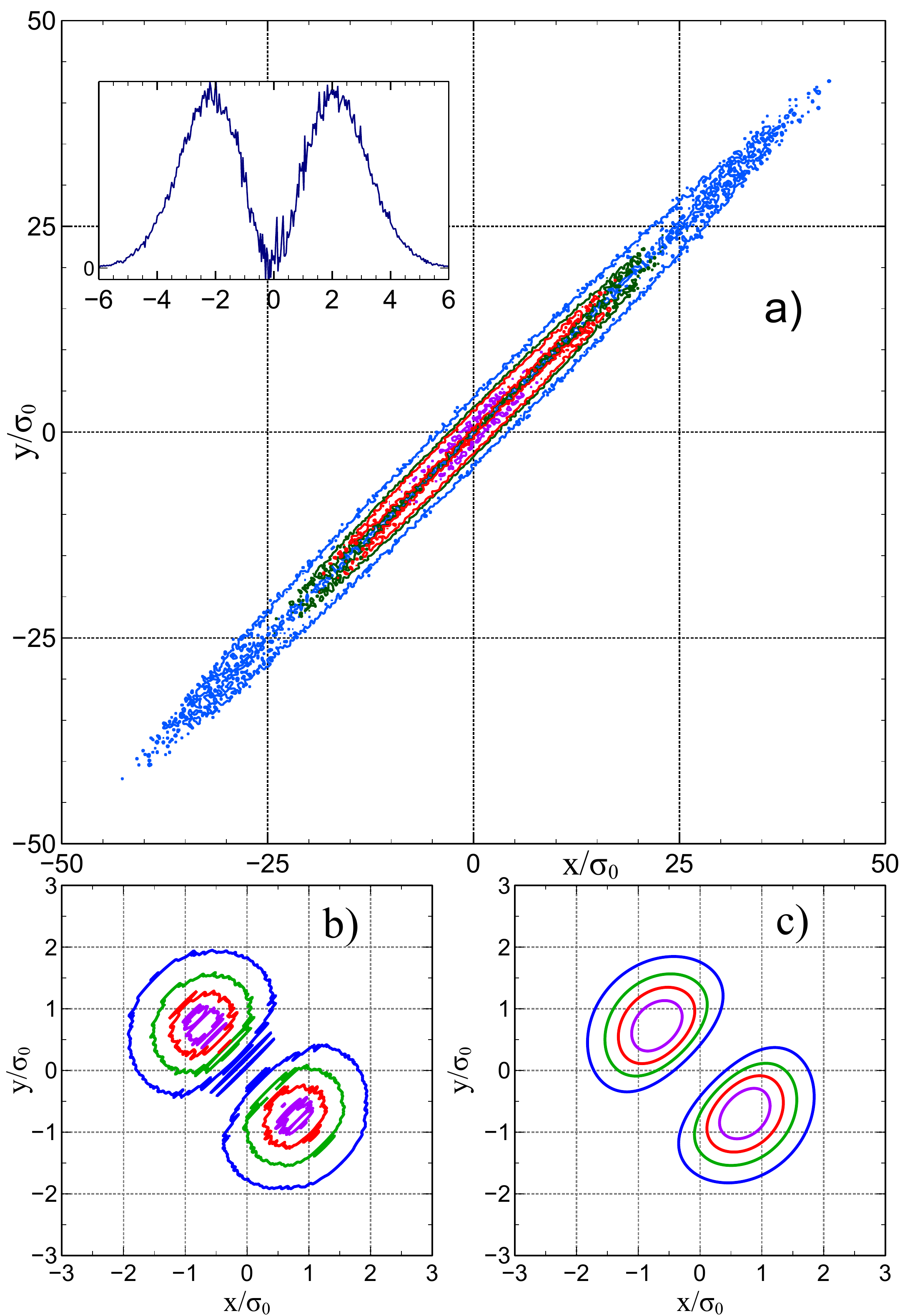}
\caption{\label{pc1c21F} Color online. Contour map of the joint probability distribution $P(x,y)$ for the two detectors outcomes for the  $\vert 1 \rangle$ Fock state deduced using Eq. (\ref{conunsolomu}) from the  maps of the vacuum and a PRCS with $\mu=0.25$ according. a) Joint probability distribution with the LO used in the experiment directly obtained from the data in Figs \ref{pc1c2}a and \ref{pc1c2}b. Inset: marginal distribution projected on the $y=-x$ diagonal. b) Probability distribution obtained from the maps corresponding to an ideal LO (Figs \ref{pc1c2}c and \ref{pc1c2}d). c) Theoretical value. (The contour levels colors are arbitrary).} 
\end{figure}

Figure \ref{pc1c21F}a shows the reconstructed single-photon joint probability distribution for the two detector outputs obtained from the data in Figs. \ref{pc1c2}a and \ref{pc1c2}b using Eq. (\ref{conunsolomu}). The classical correlation introduced by the excess laser amplitude noise is clearly visible in the stretching of the distribution along the main diagonal. However the nonclassical nature of the state is still visible in the rapid oscillations present along the $x=-y$ diagonal. These are  better observed on the marginal distribution along the  $x=-y$ axis shown in the inset. This distribution closely approaches the expected distribution for $n=1$ harmonic oscillator eigenfunction for which Vogel's criterion for nonclassicality  is verified  \cite{Vogel00,Lvovsky02}. 

The joint probability $P(x,y)$  for the $n=1$ Fock state corresponding to an ideal local oscillator can be obtained using Eq.  (\ref{conunsolomu}) and the data in Figs.  \ref{pc1c2}c and \ref{pc1c2}d; it is shown in Fig. \ref{pc1c21F}b. While affected by noise, the obtained joint distribution for the single-photon state preserves the main features of the theoretical distribution plotted in Fig.  \ref{pc1c21F}c. The quality of the experimentally determined joint distribution can be addressed using as a figure of merit the overlap $C\equiv\int P(x,y)P_{th}(x,y)dxdy/[\int P_0(x,y)^{2}dxdy\int P_{th}(x,y)^{2}dxdy]^{1/2}$ with the theoretical distribution $P_{th}(x,y)$. The data presented in Fig. \ref{pc1c21F}b correspond to $C=0.994$. \\

We consider now the experimental determination of the probability distribution $w(M)$ of the product $M=\delta I_1 \delta I_2$ of the photodetector fluctuations outputs. The experimentally observed statistical distributions of M for the vacuum state and for PRCSs with small values of $\mu$ ($\mu \lesssim 100$), are virtually indistinguishable   and largely dominated by the strong classical correlation introduced by the laser excess noise. A very small fraction of the total samples record correspond to negative values of $M$. In consequence, the method described above for the reconstruction of the single-photon statistics from that of the vacuum and a set of PRCSs is difficult to apply directly to the experimental distributions $w(M)$  due to the small statistical significance of the negative values of $M$.

\begin{figure}[h!]
\centering\includegraphics[width=0.8\linewidth]{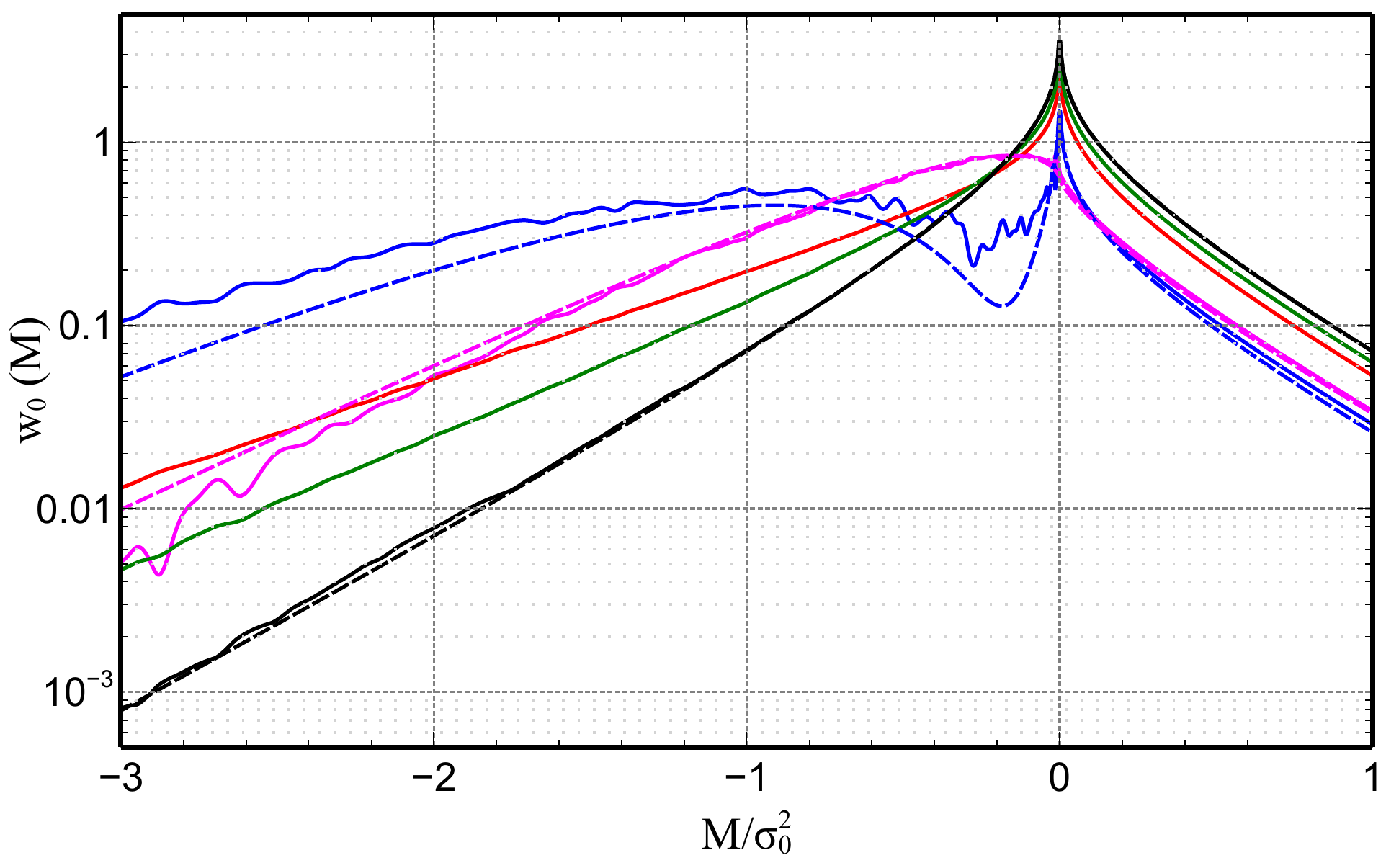}
\caption{Color online. Probability density $w_0(M)$  of the product of the photodetector fluctuations. Solid lines: Experimental probability distributions obtained using Eq. (\ref{w}) for vacuum (black), PRCS with $\mu_1=0.27$ (green), PRCS with $\mu_2=0.62$ (red). Distributions deduced from Eqs. (\ref{masdeunmu}) for Fock states $\vert 1\rangle$ (magenta) and $\vert 2\rangle$ (blue). Dashed lines: theoretical predictions for Fock states $\vert 0\rangle$, $\vert 1\rangle$  and $\vert 2\rangle$  \cite{kuhn18}.} \label{WMfig}
\end{figure} 

This difficulty can be overcome by computing first from the experimental data the  distribution $w_0(M)$ corresponding to an ideal LO using  Eq. (\ref{w}). Then the obtained distributions for the ideal LO can be combined using Eqs. (\ref{conunsolomu}) or (\ref{masdeunmu}) to obtain the probability density $w_0(M)$ for Fock states.

Figure \ref{WMfig} shows the plots of the probability densities $w_0(M)$ obtained through Eq. (\ref{w}) from the experimental observations for the vacuum and two PRCSs with $\mu_1=0.27$ and $\mu_2=0.62$. As expected the distribution is asymmetric for  nonzero mean value PRCSs  with larger weight of the negative values of $M$ as a consequence of energy conservation. The asymmetry increases as $\mu$ increases. Also shown in Fig. \ref{WMfig} are the probability densities  for Fock states $\vert 1\rangle$ and $\vert 2\rangle$ obtained from the densities for the vacuum and the PRCSs through Eqs. (\ref{masdeunmu}). The dashed lines are the theoretical probability densities  for the two considered Fock states \cite{kuhn18}. As observed, the experimentally obtained probability density for the $\vert 1\rangle$ state closely approaches the theoretical prediction. The reconstructed probability distribution is less accurate for the $\vert 2\rangle$ state although the most interesting features such as the occurrence of 
oscillations is clearly displayed. It is worth noting here that the probability distributions obtained through Eqs. \ref{conunsolomu} and \ref{masdeunmu} are only approximately normalized \cite{Valente17}.

Using as a figure of merit the overlap $D\equiv\int W(M)W_{th}(M)dM/[\int W(M)^{2}dM\int W_{th}(M)^{2}dM]^{1/2}$ of the experimental probability distribution with the theoretical prediction, the data presented in Fig. \ref{WMfig} for states $\vert 1\rangle$ and $\vert 2\rangle$ correspond to $D=0.999$ and $D=0.984$ respectively. In principle, the accuracy of the reconstructed probability densities can be arbitrarily improved through the use of a larger set of values for $\mu$.\\

In the preceding paragraphs the detector efficiency was considered ideal. Let us now briefly discuss this assumption. Provided that the detectors quantum efficiencies are the same, their role, as well as that of a possible mode-mismatch between the signal and the LO, is equivalent to an overall attenuation of the PRCS before entering the BHD setup. The two situations corresponding on one hand to ideal photodetectors and a PRCS with mean photon number $\mu$ or, on the other hand, to inefficient detectors with efficiency $\eta < 1$ and a PRCS with mean photon number $\mu/\eta$ will result in identical photodetection statistics. In consequence, PRCSs are not suitable for detector efficiency calibration \cite{Cooper14} unless the mean photon number is known through an independent determination.

In our experiment the detectors are similar photodiodes from the same manufacturing series for which similar efficiencies are reasonably expected. The values of $\mu$ assigned by the fitting procedure to the PRCS and used in the preceding sections already incorporate the effects of detector inefficiency and mode-mismatch. They correspond to PRCS that produce the observed output signals when acting on \emph{ideal} photodetectors. It is nevertheless  straightforward to simulate the response that would result from inefficient photodetectors with a given (common) efficiency $\eta$ ($\eta < 1$). For this it is enough to overestimate the value of PRCS mean photon-number by replacing $\mu$ by $\mu/\eta$ when using Eqs. \ref{conunsolomu} and \ref{masdeunmu}.

\section{Conclusions}

In summary, we were able to extract from experimental data obtained in a balanced homodyne detection setup using a LO oscillator with excess amplitude noise, the ideal full two-detector statistics corresponding to  a LO in a shot noise limited coherent state. The method was applied to phase randomized coherent states from which the statistics of Fock states was derived. This allowed us to present the first experimental recording of the statistical distribution of the product of outcomes of the two detectors outputs for Fock states recently predicted in \cite{kuhn18}.

% two-detector  of the first experimental test This allowed us to present the first  
%In summary, we were able to obtain from data originating in a BHD setup using a noisy LO the complete ideal statistics of the two photodetector outputs. Joint probability distributions were observed for phase randomized coherent states and Fock states. In addition, the first experimental measurement of the two detector correlation statistics of Fock states recently calculated in \cite{kuhn18} was presented. 

To obtain the ideal statistical distributions from the apparatus including a noisy LO, two sets of measurements are needed: a first set for the determination of the statistical distribution of the field quadrature and a second set to determine the variance of the vacuum fluctuations. The resources cost of this second measurement is quite modest since a smaller data set, compared to the one required to establish the quadrature probability density profile,  will suffice to accurately estimate the variance of vacuum fluctuation. 

This reflects the fundamental fact that in terms of information contents, the additional information present in the full two photodetector statistics as compared to just the field quadrature distribution amounts to a real number corresponding to the vacuum fluctuation variance [see  Eqs. (\ref{newjoint}) and (\ref{w})]. It is interesting to notice that this additional information requirement is the same that is needed to establish nonclassicality directly from the field quadrature distribution using Vogel's criterion \cite{Vogel00,Lvovsky02}. \\

%\begin{acknowledgments}
This work was supported by ANII, CSIC and PEDECIBA (Uruguayan agencies).
%\end{acknowledgments}

%\bibliography{wMAA}

\begin{thebibliography}{29}
\bibitem [{noe()}]{noentanglement}%
  \BibitemOpen
  \href@noop {} {}\bibinfo {note} {It is assumed that there is no entanglement
  between the states of modes $a$ and $b$.}\BibitemShut {Stop}%
\end{thebibliography}


\end{document}